\documentstyle[prl,aps]{revtex}

\def\textbf#1{{\bf #1}}

\def\be{\begin{equation}}
\def\ee{\end{equation}}
\def\ben{\begin{eqnarray}}
\def\een{\end{eqnarray}}
\def\eea{\end{array}}
\def\bea{\begin{array}}
\newcommand{\bei}{\begin{itemize}}
\newcommand{\eei}{\end{itemize}}

\begin{document}
\draft

\title{Erratum: Asymptotic entanglement manipulations can be genuinely 
irreversible. [Phys. Rev. Lett. {\bf 84}, 4260 (2000)] }

\author{Micha\l{} Horodecki$^{1}$,
Pawe\l{} Horodecki$^{2}$ and Ryszard Horodecki$^{1}$}

\address{$^1$ Institute of Theoretical Physics and Astrophysics,
University of Gda\'nsk, 80--952 Gda\'nsk, Poland,\\
$^2$Faculty of Applied Physics and Mathematics,
Technical University of Gda\'nsk, 80--952 Gda\'nsk, Poland
}

\maketitle

\begin{abstract}
\end{abstract}


The presented proof in Ref. \cite{irrev} that $E_D<E_f^\infty$ 
(irreversibility) for some Werner states, was based on an invalid 
lemma of Ref. \cite{Rains} (cf. erratum \cite{Rainse}) (p. 3 of the 
Ref \cite{irrev}) saying that $E_{PT}$
is additive for Werner states. Therefore our proof of irreversibility 
is incorrect. However, the irreversibility holds, as shown recently
in Ref. \cite{VC}. The authors considered some family of 
bound entangled states (hence having $E_D=0$) and 
showed that $E_f^\infty$ for those states is nonzero. 

Other results of the our paper remain valid 
as they do not make use of the mentioned lemma.
They are:
\bei
\item Proof that $E_n=\log||\varrho^{PT}||$ is upper bound for 
distillable entanglement
(Lemma 2 of Ref. \cite{irrev}). An independent proof of this fact was 
found earlier by Werner, Benasque, 1998 [private communication], see Ref \cite{VW}. 
\item Proof that $E_n$ does not increase under trace-preserving 
PPT superoperators (Appendix of Ref. \cite{irrev}).
\item Calculation of value of $E_{PT}$ for Werner states (eq. (16) of 
Ref. \cite{irrev}). 
\item Calculation of the measure $E_n$ for isotropic state 
and Werner states (eq. (13) and (15) of Ref. \cite{irrev}).
\eei

\end{document}